\documentclass[pre,reprint,twocolumn,superscriptaddress]{revtex4}

\usepackage{graphicx}
\usepackage[caption=false]{subfig}
\usepackage{amsmath}
\usepackage{amssymb}
\usepackage{color}
\usepackage{siunitx}
\usepackage{dcolumn}
\usepackage{cancel}

\begin{document}
%%%

%%%%%General information

\title{Stochastic motion under nonlinear friction representing shear thinning}

%%% Authors
\author{Theo Lequy}
\email{theo.lequy@posteo.de}
\affiliation{Eidgen{\"o}ssische Technische Hochschule Z\"urich, R{\"a}mistrasse 101, 8092 Z\"urich, Switzerland}

\author{Andreas M. Menzel}
\email{a.menzel@ovgu.de}
\affiliation{Institut f{\"u}r Physik, 
Otto-von-Guericke-Universit\"at Magdeburg, Universit\"atsplatz 2, 39106 Magdeburg, Germany}

%%% Date
\date{\today}

\begin{abstract}
We study stochastic motion under a nonlinear frictional force that levels off with increasing velocity. 
Specifically, our frictional force is of the so-called Coulomb-tanh type. At small speed, it increases approximately linearly with velocity, while at large speed it approaches a constant magnitude, similarly to solid (dry, Coulomb) friction. In one spatial dimension, a formal analogy between the associated Fokker-Planck equation and the Schr\"odinger equation for a quantum-mechanical oscillator in a nonharmonic P\"oschl-Teller potential is revealed. Then, the stationary velocity statistics can be treated analytically. From such analytical considerations, we determine associated diffusion coefficients, which we confirm by agent-based simulations. Moreover, from such simulations and from numerically solving the associated Fokker-Planck equation, we find that the spatial distribution function, starting from an initial Gaussian shape, develops tails that appear exponential at intermediate time scales. At small magnitudes of stochastic driving, the velocity distribution resembles the case of linear friction, while at large magnitudes it rather approaches the case of solid (dry, Coulomb) friction.  The same is true for diffusion coefficients. In a certain sense, thus interpolating between the two extreme cases of linear friction and solid (dry, Coulomb) friction, our approach should be useful to describe several cases of practical relevance. For instance, a reduced increase in friction with increasing relative speed is typical of shear-thinning behavior. Therefore, driven motion in shear-thinning environments is one specific example to which our description may be applied. 
\end{abstract}

%\begin{document}

\maketitle
%\tableofcontents

\section{Introduction}

Addressing the dynamics of an object that is subject to both linear friction with its environment and a white stochastic driving force of Gaussian distribution %and vanishing mean 
is a classical textbook example \cite{risken1996fokker}. The procedure of deriving the corresponding velocity and spatial displacement statistics by crossing from the Langevin dynamics to the associated Fokker-Planck equation is well established. In that case, the steady-state velocity distribution is of Gaussian form. Moreover, an initially single-peaked spatial probability distribution of finding the object at a certain location likewise adopts Gaussian shape over time, shrinks in height, and broadens in width. 

To describe the stochastically driven motion of objects that slide on a solid substrate, it is common to include the so-called solid or dry friction, sometimes also referred to as Coulomb friction \cite{persson2013sliding}. This frictional force at nonvanishing speed is of constant magnitude and always oriented against the current velocity direction. The associated steady-state velocity distribution features a cusp at vanishing speed and, in the absence of linear friction, is of purely exponential character \cite{degennes2005brownian, hayakawa2005langevin, touchette2010brownian, menzel2011effect, touchette2012exact, das2017single}. Initially single-peaked spatial distribution functions develop intermediate non-Gaussian shapes and tails, while the mean-squared displacement apparently still increases linearly in time \cite{menzel2011effect}. Experimental investigations on a broad variety of different systems provide related observations \cite{wang2012brownian}.

Solid (dry) friction of the Coulomb type is an extreme example. From zero velocity, it directly jumps to a constant magnitude when the object is set into motion. Similarly, when reverting the velocity direction starting from small but nonvanishing speed, this type of friction abruptly jumps from the finite initial value to the value of opposite sign but identical magnitude.

One way of loosening this peculiar property of finite jump at vanishing speed is to replace the associated functional form of friction with respect to the velocity by a hyperbolic tangent. It smoothly and steadily crosses the point of vanishing speed with finite slope. Yet, it remains bounded for increasing magnitude of velocity. This so-called Coulomb-tanh friction model has been introduced and applied before \cite{andersson2007friction, pennestri2016review, wang2021time}. Here, we illustrate a way to evaluate this friction model in the context of motion of a stochastically driven object. Our statistical approach allows for some analytical considerations including velocity statistics and diffusion coefficients, whereas the spatial statistics are evaluated numerically. 

We point out that one possible context of dynamics under Coulomb-tanh friction is, to some degree, motion in a shear-thinning environment \cite{morris2009review}. Generally, for relative motion under shear thinning, the increase in friction decreases with increasing relative speed. At small magnitudes of velocity, Coulomb-tanh friction increases approximately linearly with speed. This increase is reduced by rising speed, which, for example, can be considered as the mentioned characteristic feature of shear-thinning behavior. If regarded in this context and along these lines, we may interpret the mathematical form of solid (dry, Coulomb) friction as an extreme limit. Contrarily but in a similar spirit, the statistics of stochastic motion in a shear-thickening environment has also been evaluated in the context of a nonlinear friction term \cite{menzel2015velocity}. 

Similarly to previous initial treatments of stochastic motion under solid (dry) friction of the Coulomb type \cite{degennes2005brownian, touchette2010brownian, menzel2011effect}, here we confine ourselves to motion in one spatial dimension. This allows for several analytical considerations. 
Our presentation is structured as follows. Next, in Sec.~\ref{sec_langevin}, we present the underlying stochastic equations of Langevin and Fokker-Planck types. Particularly, we introduce an appropriate ``velocity potential'' in the Fokker-Planck equation that reproduces the Coulomb-tanh friction term in the Langevin equation. As a consequence, we find an analytical expression for the velocity distribution in Sec.~\ref{sec_velocity} and can analytically approach its dynamics in Sec.~\ref{sec_timev}. On this basis, an expression of closed integral form is obtained for the diffusion coefficient in Sec.~\ref{sec_diff_ceoff}. The dynamics of the decaying spatial distribution is illustrated by numerical solution in Sec.~\ref{sec_spatial}. We conclude in Sec.~\ref{sec_concl} by presenting an outlook on possible future investigations. Some technical details are shifted to the four appendices. 

\begin{figure}
    \centering
    \includegraphics[width=0.99\linewidth]{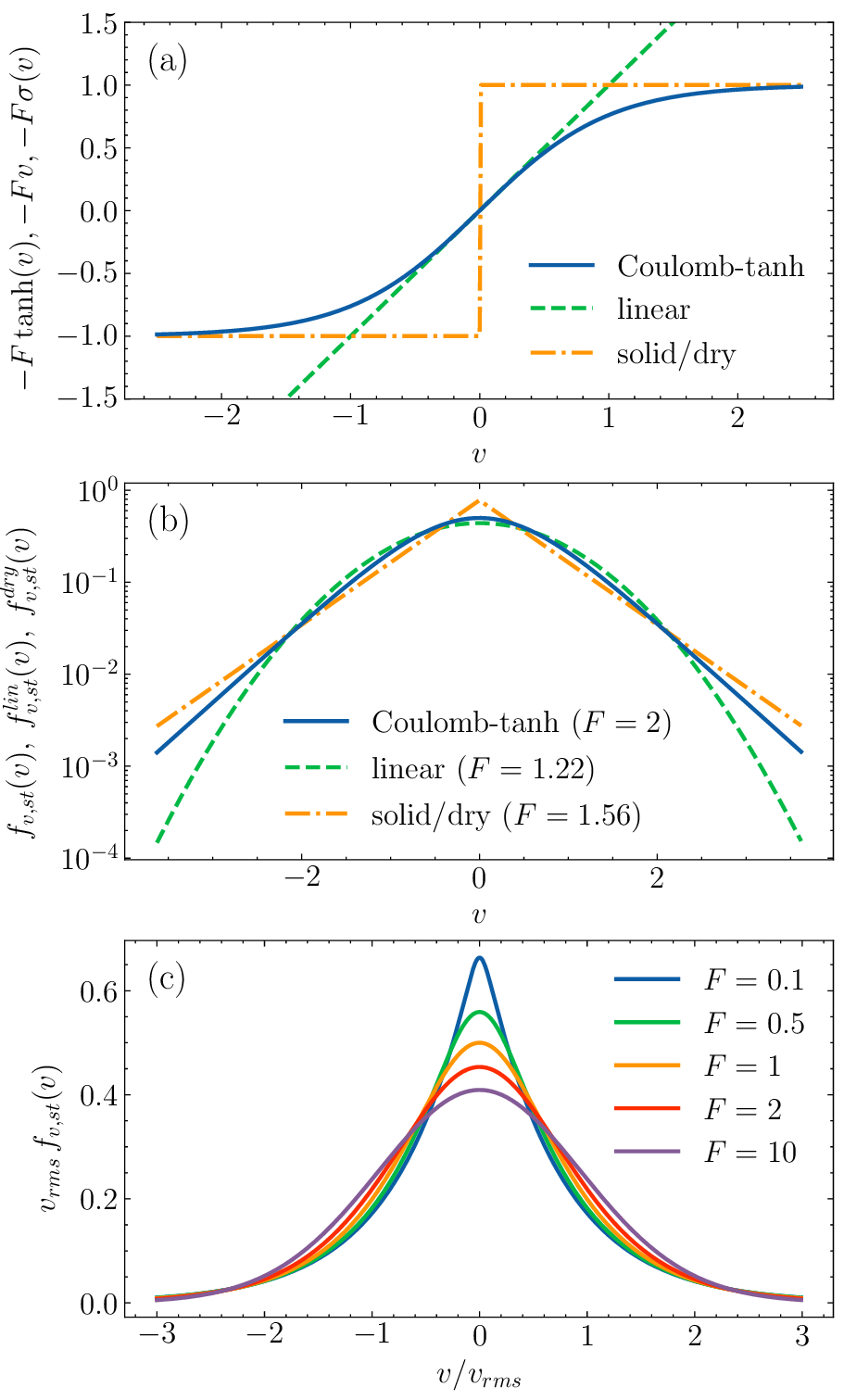}
    \caption{Friction of the Coulomb-tanh type $-F\tanh(v)$ (blue, solid) in comparison to linear friction $-Fv$ (green, dashed) and solid (dry, Coulomb) friction $-F\sigma(v)$ (orange, dash-dotted) as well as stationary velocity distributions $f_{v,st}(v)$. 
    (a) At small magnitudes of velocity $|v|$, the friction of the Coulomb-tanh type increases linearly with $v$, while, at large $|v|$, it levels off and approaches the limit of solid (dry, Coulomb) friction. 
    (b) Stationary velocity distributions $f_{v,st}(v)$, $f_{v,st}^{lin}(v)$, and $f_{v,st}^{dry}(v)$ [see Eqs.~(\ref{eq:fvst}), (\ref{eq:fvst_linear}), and (\ref{eq:fvst_solid_dry}), respectively] when compared to each other.  The friction parameters $F=2$ for Coulomb-tanh friction, $F=1.22$ for linear friction, and $F=1.56$ for solid (dry, Coulomb) friction were adjusted to render the mean-squared velocity equal. They imply an equal effective temperature of $T_v(2)\approx T_v^{lin}(1.22)\approx T_v^{dry}(1.56)\approx 0.82$ as defined by Eqs.~(\ref{eq:T_dimless}), (\ref{eq:Tv_lin}), and (\ref{eq:Tv_dry}). 
    (c) Stationary velocity distribution $f_{v,st}(v)$ for Coulomb-tanh friction, see Eqs.~(\ref{eq:fvst}), at different force parameters $F$. For better comparison, we rescale the velocity and the distribution function using the root-mean-squared velocity $v_{rms}$ in each case. 
    \label{fig:overview_Ffric_fvst}}
\end{figure}

\section{Statistical equations}\label{sec_langevin}
Theoretically, the stochastic motion of an object in one spatial dimension under Coulomb-tanh friction is described by the following coupled differential equations of the Langevin type,
\begin{subequations} \label{eq:langevin}
\begin{eqnarray} \label{eq:langevin_v}
     m\frac{d\Tilde{v}}{d\Tilde{t}} &=& {}- \Tilde{F}\tanh\!\left(\frac{\Tilde{v}}{v_0}\right) + \Tilde{\gamma} (\Tilde{t}),\\
    \frac{d\Tilde{x}}{d\Tilde{t}} &=& \Tilde{v}.
    \label{eq:langevin_x}
\end{eqnarray}
\end{subequations}
In these equations, $\Tilde{x}$, $\Tilde{v}$, and $\Tilde{t}$ denote the still dimensionful spatial position, velocity, and time, respectively. $m$ is the mass of the object. 
The first term $-\Tilde{F}\tanh(\Tilde{v}/v_0)$ on the right-hand side of Eq.~(\ref{eq:langevin_v}) introduces the mathematical expression of the specific frictional force of the Coulomb-tanh type that we mentioned above. The parameter $v_0$ sets the slope at infinitely small speed, which is $\Tilde{F}/v_0$. 
Moreover, the last term of Eq.~(\ref{eq:langevin_v}), $\Tilde{\gamma}(\Tilde{t})$, includes the stochastic force. We assume it to be $\delta$-correlated in time, white, and of Gaussian type, $\langle \Tilde{\gamma}(\Tilde{t}),\Tilde{\gamma}(\Tilde{t}') \rangle = 2K\delta(\Tilde{t}-\Tilde{t}')$. Here, $K$ sets the strength and $\delta$ denotes the Dirac delta function. Furthermore, we require  $\langle \Tilde{\gamma}(\Tilde{t})\rangle = 0$.

We remark in this context that, as is apparent from relating the strength of the stochastic force to a constant value $K$, the situation that we are addressing is not an equilibrium one. In the latter case, a fluctuation-dissipation theorem would apply that balances the input of energy to our object due to stochastic driving caused by thermal fluctuations with the loss in energy due to friction and dissipation. In general, its form under nonlinear friction differs from the fluctuation-dissipation theorem for linear friction \cite{kubo1991nonequilibrium}.

In contrast to such an equilibrium situation, here we rather think of a scenario in which the stochastic driving is caused by a defined external mechanism. %, the strength of which on average is set constant. 
For example, related earlier works by de Gennes \cite{degennes2005brownian} and Chaudhury \textit{et al.}\ \cite{goohpattader2009experimental,goohpattader2010diffusive} assume and experimentally investigate an object on a vibrated substrate. To this situation, they apply the described formalism setting the ensemble-averaged strength of the stochastic force constant. The stochastic force there is determined by the externally controlled magnitude and type of applied vibration. In their case, dry (solid) friction results between the substrate and the object, which is modeled in the calculations by a nonlinear friction term of the Coulomb type.

In our case, we likewise assume a driving force of constant averaged strength and therefore a similar setting. Yet, we consider a modified type of friction between the object and the surface. For instance, a layer of polymer solution or melt may be deposited onto the vibrated substrate. Thus, the vibrated object slides on this layer and shears it between itself and the substrate. This shear may cause nonlinear friction that increases less substantially with speed the higher the speed of the sliding object becomes. The nonlinearity of the friction then results from the shear-thinning behavior of the deposited layer between the sliding object and the vibrated substrate. This trend of behavior is modeled in our approach by the nonlinear Coulomb-tanh friction force. Confinement to one-dimensional motion can be imposed in experiments using appropriate spatial constrictions. In that context, toroidal cavities of diameters that are large compared to the dimension of the vibrated objects could be used to mimic one-dimensional motion under periodic boundary conditions \cite{volfson2004anisotropy}.

Next, we turn to dimensionless quantities by rescaling $v = \Tilde{v}/v_0$, $t = K\Tilde{t}/m^2v_0^2 $, $F =mv_0\Tilde{F}/K$, and $x = K\Tilde{x}/m^2v_0^3 $. 
Now, the frictional force of the Coulomb-tanh type simply reads $-F\tanh(v)$. 
Its magnitude is bounded by the parameter $F>0$ in the limit of large speed. There, the frictional force levels off and approaches the magnitude of solid (dry, Coulomb) friction $-F\sigma(v)$ [see Fig.~\ref{fig:overview_Ffric_fvst}(a)], where $\sigma(v)$ with $\sigma(0)=0$ denotes the sign function. At small speed, we recover standard linear friction $-Fv$.

Finally, we introduce the probability density in phase space, $f(x,v,t)$, of finding an object at time $t$ at position $x$ with velocity $v$. Its dynamic evolution is described by a continuum equation of the Fokker-Planck type. It is derived via standard procedures \cite{risken1996fokker, zwanzig2001nonequilibrium} and reads 
\begin{equation}\label{eq:FP}
    \partial_t f(x,v,t) = \Bigl\{\!{} - v \partial_x + \partial_v F\tanh{v} + \partial_v^2\Bigr\} \,f(x,v,t).
\end{equation}

\section{Steady-state velocity distribution}\label{sec_velocity}

We find the differential equation for the velocity distribution $f_v(v,t) = \int_{-\infty}^{\infty}f(x,v,t) dx$ by integrating out the spatial position $x$ from Eq.~(\ref{eq:FP}), resulting in
\begin{equation} \label{eq_FP_v}
    \partial_t f_v(v,t) = \partial_v\Bigl\{ F\tanh{v} +\partial_v\Bigr\} f_v(v,t).
\end{equation} 
From there, we obtain the normalized stationary velocity distribution via $\partial_t f_v(v,t) =0$,
\begin{subequations}
\label{eq:fvst}
\begin{equation}\label{eq:fvst_without_coefficient}
    f_{v,st}(v) = \frac{\cosh^{-F}(v)}{\int_{-\infty}^{\infty} \cosh^{-F}(v)\;dv}. 
\end{equation}
This expression can be rewritten as 
\begin{eqnarray} \label{eq:fvst_with_coefficient}
f_{v,st}(v)&=&\frac{F!F}{2^{F+1}(F/2)!^2} \cosh^{-F}(v)\nonumber \\  &=& \frac{\Gamma\!\left(\frac{F+1}{2}\right)}{\sqrt{\pi}\,\Gamma\!\left(\frac{F}{2}\right)} \cosh^{-F}(v),
\end{eqnarray}
\end{subequations}
see Appendix~\ref{sec_appA}, where $\Gamma$ denotes the Gamma function.

Examining this distribution and comparing it to the cases of linear and purely solid (dry, Coulomb) friction, it generally shows an intermediate shape, see Fig.~\ref{fig:overview_Ffric_fvst}(b). At small speed, it resembles a Gaussian distribution due to the approximate linearity of the frictional force in this regime. At large speed, the tails appear exponential as in the case of solid (dry, Coulomb) friction, which is related to the leveling off of the frictional force. 

Additional dependence of the shape of $f_{v,st}(v)$ results from the parameter $F>0$. As we can see in Fig.~\ref{fig:overview_Ffric_fvst}(c), with decreasing values of $F$, the Gaussian regime becomes very narrow and the distribution approaches the character for solid (dry, Coulomb) friction, which features a cusp at $v=0$. The exponential tails dominate. Contrarily,  for increasing values of $F$, the central Gaussian character broadens and the exponential tails are pushed further out.
In Fig.~\ref{fig:overview_Ffric_fvst}(c), we rescale the velocity $v$ by its root-mean-squared value $v_{rms}$, to better fit the different curves into one diagram. Accordingly, we multiply $f_{v,st}(v)$ by $v_{rms}$ to preserve normalization.

For later reference, we define a temperaturelike variable $T_v$ for the object as its mean-squared velocity, $T_v := \langle v^2 \rangle_{st}$, where $\langle ... \rangle_{st}=\int_{-\infty}^{\infty}...f_{v,st}(v)dv$; see Eqs.~(\ref{eq:fvst}). We obtain
%\begin{subequations} \label{eq:T}
\begin{equation} \label{eq:T_dimless}
    T_v(F)= \frac{1}{2}\psi^{(1)}\left(\frac{F}{2}\right),
\end{equation}
where $\psi^{(1)}(z) = d^2 \ln\Gamma(z)/dz^2$ is the second polygamma function, also called the trigamma function; see Appendix~\ref{sec_appB}. 

At low temperatures, the object mainly probes the regime of small speed. There, friction depends approximately linearly on velocity $v$. We obtain the corresponding frictional form $-Fv$ in the limit of small magnitudes of velocity from our friction of the Coulomb-tanh type $-F\tanh(v)$. Again, $F$ parameterizes the strength of the frictional force. 
From the appropriate replacement of $-F\tanh(v)$ by $-Fv$ in Eq.~(\ref{eq:FP}), we obtain the corresponding stationary velocity distribution $f_{v,st}^{lin}(v)$. It is of Gaussian shape and reads
\begin{equation}\label{eq:fvst_linear}
    f_{v,st}^{lin}(v) = \sqrt{\frac{F}{2\pi}}\exp\left(-\frac{F v^2}{2}\right),
\end{equation}
see Fig.~\ref{fig:overview_Ffric_fvst}(b). Based on this stationary velocity distribution, we find the temperature
\begin{equation}\label{eq:Tv_lin}
    T_v^{lin}(F)=\langle v^2\rangle_{st}^{lin}=\frac{1}{F}. 
\end{equation}

 Contrarily, at high temperatures, large values of the speed of the object dominate. In this limit of large magnitudes of velocity $v$, the frictional form of the Coulomb-tanh type turns into $-F\sigma(v)$, associated with the leveling off in the frictional force. Here, $\sigma(v)$ denotes the sign function, where $\sigma(0)=0$. This functional form is identical to the case of solid (dry, Coulomb) friction.  
 Again, substituting the frictional term in Eq.~(\ref{eq:FP}) accordingly, we find the associated stationary velocity distribution $f_{v,st}^{dry}(v)$. 
 It is of exponential shape, 
     \begin{equation} \label{eq:fvst_solid_dry} 
         f_{v,st}^{dry}(v) = \frac{F}{2} \exp(-F|v|).
     \end{equation}
 Figure~\ref{fig:overview_Ffric_fvst}(b) illustrates the corresponding result. Using this stationary velocity distribution, the associated temperature reads
\begin{equation}\label{eq:Tv_dry}
    T_v^{dry}(F)=\langle v^2\rangle_{st}^{dry}=\frac{2}{F^2}. 
\end{equation}

\begin{figure}
    \centering
    \includegraphics[width=0.99\linewidth]{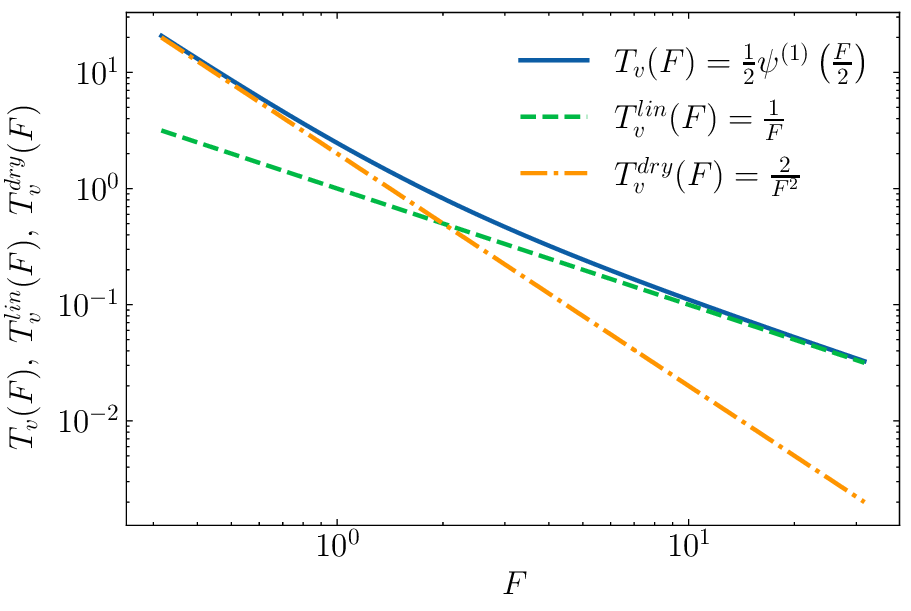}
    \caption{Effective temperature $T_v(F)$ as a function of the strength $F$ for Coulomb-tanh friction $-F\tanh{v}$ (blue, solid), when compared to temperatures $T_v^{lin}(F)$ for linear friction $-Fv$  (green, dashed) and $T_v^{dry}(F)$ for solid (dry, Coulomb) friction $-F\sigma(v)$  (orange, dash-dotted). The asymptotic behavior at small and large $F$ is well reproduced. 
    % \am{Bitte überall $T_v(F)$ (mit $F$), linear zuerst (hatten wir überall so), dry und linear in den Superskript.}
    \label{fig:temp_comp}}
\end{figure}

\begin{figure*}
    \centering
    \includegraphics[width=\textwidth]{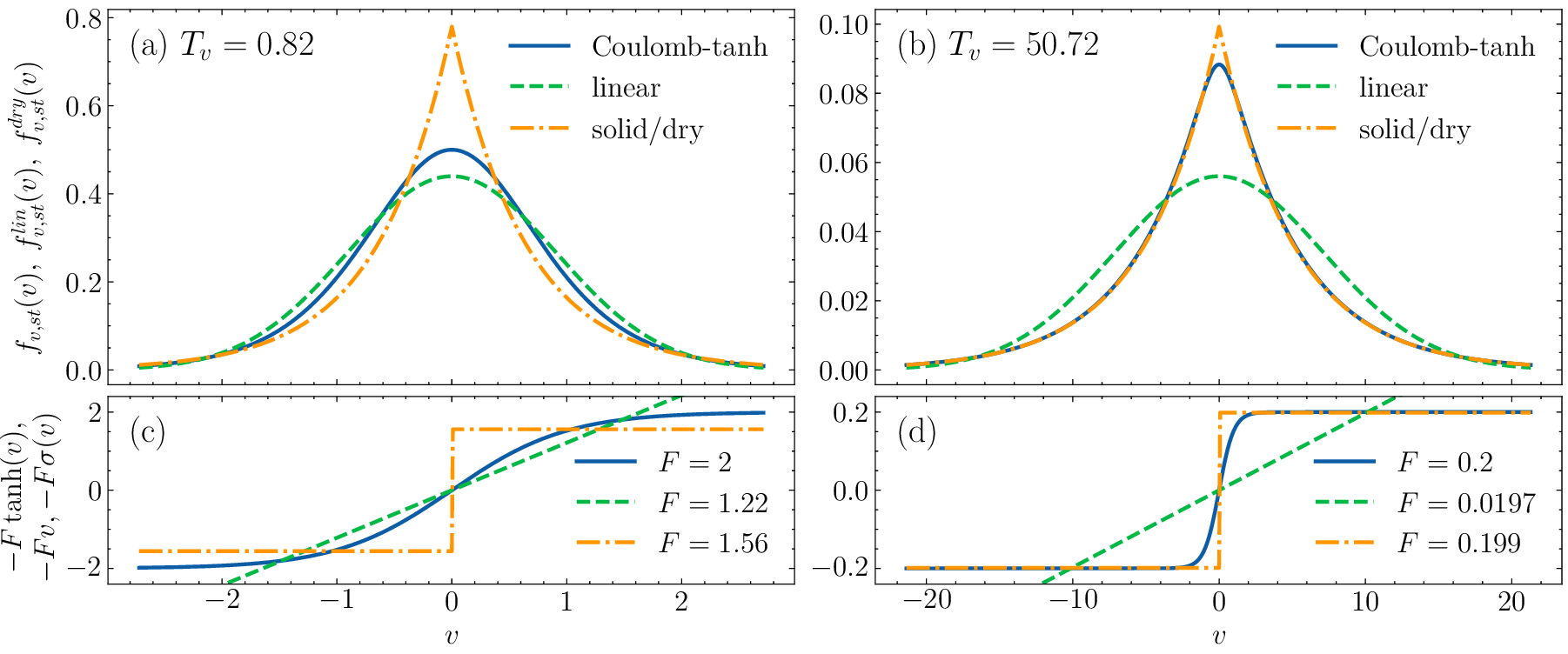}
    \caption{(a),(b) Stationary velocity distributions $f_{v,st}(v)$, $f_{v,st}^{lin}(v)$, and $f_{v,st}^{dry}(v)$ [see Eqs.~(\ref{eq:fvst}), (\ref{eq:fvst_linear}), and (\ref{eq:fvst_solid_dry}), respectively] and (c),(d) functional forms of the frictional force for Coulomb-tanh friction $-F\tanh(v)$ (blue, solid), linear friction $-Fv$ (green, dashed), and solid (dry, Coulomb) friction $-F\sigma(v)$ (orange, dash-dotted), at different effective temperatures (a,c) $T_v(2)\approx T_v^{lin}(1.22)\approx T_v^{dry}(1.56)\approx 0.82$ and (b,d) $T_v(0.2)\approx T_v^{lin}(0.0197)\approx T_v^{dry}(0.199) \approx 50.72$; see Eqs.~(\ref{eq:T_dimless}), (\ref{eq:Tv_lin}), and (\ref{eq:Tv_dry}). The values of $F$ for the different frictional forces were adjusted as listed in (c) and (d) to render the temperatures for all frictional cases equal in (a),(c) and (b),(d). 
    At small temperature (a),(c), the stationary velocity distribution $f_{v,st}(v)$ for the Coulomb-tanh type resembles the Gaussian shape of $f_{v,st}^{lin}(v)$ for linear friction. 
    Conversely, at large temperature (b),(d), it approaches the shape of $f_{v,st}^{dry}(v)$ for solid (dry, Coulomb) friction of dominating exponential character.
    \label{fig:fvst_Ffric_comp}}
\end{figure*}

The trigamma function $\psi^{(1)}(F/2)$ in Eq.~(\ref{eq:T_dimless}) for positive $F$ decreases with increasing force parameter $F$. Therefore, from Eq.~(\ref{eq:T_dimless}), we infer that large values of $F \gg 1$ are related to small temperatures $T_v(F)$. In this regime, the object mainly attains small speeds and therefore predominantly probes the regime of linear friction $-Fv$. From the asymptotic behavior of the trigamma function, we observe in this case that for $F \gg 1$, the temperature decreases with increasing $F$ as $T_v(F) \approx F^{-1}$. This is the same functional form as for the temperature $T_v^{lin}(F)$ in the case of linear friction; see Eq.~(\ref{eq:Tv_lin}).
Conversely, small values of $F \ll 1$ are related to high temperatures $T_v(F)$. Then, the object is mostly found at large speeds, where friction resembles the solid (dry, Coulomb) type $-F\sigma(v)$. 
From the asymptotic behavior of the trigamma function, we find that for these values of $F \ll 1$, temperature decreases as $T_v(F)\approx 2F^{-2}$. This recovers the functional dependence of the temperature  $T_v^{dry}(F)$ in the case of solid (dry, Coulomb) friction; see Eq.~(\ref{eq:Tv_dry}). The asymptotic behavior is illustrated in Fig.~\ref{fig:temp_comp}.
We compare the stationary velocity distributions for the different types of friction to each other at different temperatures in Fig.~\ref{fig:fvst_Ffric_comp}.

\section{Time evolution of velocity statistics}\label{sec_timev} 

In search of an analytical approach to Eq.~(\ref{eq_FP_v}), we turn to the eigenfunctions of the operator on its right-hand side. Multiplying the equation by $f_{v,st}^{-1/2}(v)$ from the left and using the substitution $ g_v :=f_{v,st}^{-1/2}f_v$, we eliminate the first derivative $\partial_v$ and render the operator Hermitian. The resulting equation then resembles a quantum mechanical Schr\"odinger equation,
\begin{equation}
    \partial_t g_v = \Biggl\{-\frac{F^2}{4}+ \frac{F(F+2)}{4 \cosh^2v} + \partial_v^2 \Biggr\} g_v.
\end{equation}
To address its time evolution, we introduce the ansatz $g_v(v,t) = \psi(v)\exp\{-(2E + F^2/4)t\}$ and find
\begin{equation} \label{eq:schroedinger}
    E \psi(v) = \Biggl\{-\frac{1}{2}\partial_v^2- \frac{F(F + 2)}{8\cosh^2v}  \Biggr\}\psi(v).
\end{equation}
This equation actually is formally equivalent to a time-independent quantum-mechanical Schr\"odinger equation, here in one-dimensional velocity space. $-\partial_v^2/2$ is identified with the momentum operator and the term containing $F$ with the external potential. Indeed, if we define $\lambda := F/2$, we recover the Schr\"odinger equation for the so-called P\"oschl-Teller potential $V(v) = - {\lambda(\lambda +1)}/{2 \cosh^2v}$ \cite{Pschl1933BemerkungenZQ}.

On the one hand, for positive integer values of $\lambda = l>0$, the discrete eigenvalues of this Schr\"odinger equation are $E = -m^2/2<0$ for $m = 1,2,\dots,l$ with the bound eigenfunctions being the associated Legendre polynomials $P_{l}^{m}(\tanh v)$. Using the orthogonality relation 
\begin{equation}
\int_{-1}^1 \frac{P_l^m(x) P_l^n(x)}{1-x^2}dx = \frac{(l+m)!}{m(l-m)!}\delta_{mn},    
\end{equation}
they can be normalized to
\begin{equation} 
    \psi_{m}(v) = \sqrt{\frac{m(l-m)!}{(l+m)!}} P_{l}^{m}(\tanh v).
\end{equation}
In particular, identifying $F=2l$ we find 
\begin{equation}
    \psi_{l}(v) = (-1)^l\frac{\sqrt{(2l)!\,l}}{2 ^ll!} %\sqrt{\frac{l}{(2l)!}}
    \cosh^{-l}(v) =(-1)^l\sqrt{f_{v,st}(v)}.
\end{equation}

On the other hand, for positive $E = k^2/2>0$ with $ k $ being a real number, we turn to the associated Legendre functions. They can be extended to complex orders $P_l^{ik}(\tanh v)$, which gives rise to a continuum of scattering states. 
A normalization is imposed by considering the asymptotic behavior at $v \to \pm \infty$, where the potential vanishes and the eigenfunctions turn into those of a free particle; see Appendix~\ref{sec_scatt_states}. 
In this way, we obtain corresponding eigenfunctions,
\begin{equation}\label{eq:eigenfunction_free}
    \psi_{ik}(v) = \frac{\Gamma(1-ik)}{\sqrt{2\pi}} P_l^{ik}(\tanh v).
\end{equation}

In combination, we may rewrite the operator on the right-hand side of Eq.~(\ref{eq_FP_v}) according to its spectral decomposition. Then, the equation reads
\begin{eqnarray} \label{eq_spectral_decom}
    \partial_t | f_v \rangle &=& f_{v,st}^{1/2} \biggl\{ \sum_m (m^2 -l^2)|\psi_m\rangle \langle \psi_m |  \\&& - \int_{- \infty}^{\infty}dk\;(l^2 + k^2)|\psi_{ik}\rangle \langle \psi_{ik} |\biggr\}f_{v,st}^{-1/2}|f_v\rangle. \nonumber 
\end{eqnarray}

\section{Diffusion Coefficient}\label{sec_diff_ceoff}
Through Einstein's equation, the diffusion coefficient can be calculated using the velocity autocorrelation function \cite{degennes2005brownian, menzel2022statistics}. The expression can be decomposed into the eigenfunctions and then, again identifying $F=2l$, reads 
\begin{eqnarray}\label{eq_D}
    D &=& \int_0^{\infty} \langle v(0)v(t) \rangle dt \nonumber \\ &=& \sum_{m}^{\text{odd}} \frac{\bigl| \langle \psi_l | v |\psi_m\rangle\bigr|^2}{l^2-m^2}  + \int_{-\infty}^{\infty}dk \frac{\bigl| \langle \psi_l | v |\psi_{ik} \rangle\bigr|^2}{l^2 + k^2}.
\end{eqnarray}

Here, only $\psi_m(v)$ of %integer $m$ with opposing parity with $l$ 
odd parity contribute to the sum, because $v$ is odd and $\psi_l(v)=(-1)^l\sqrt{f_{v,st}(v)} $ is even.

We confirm Eq.~(\ref{eq_D}) by comparison with the corresponding results for asymptotic values of the mean squared displacement $\langle x^2(t)\rangle$ obtained from agent-based simulations of Eqs.~(\ref{eq:langevin}).  
For this purpose, the equations of motion were discretized in time using the Euler method. Unless remarked otherwise, the time step was chosen as $dt=10^{-3}$. An ensemble of $N=10^5$ objects was considered. As initial conditions we randomly sampled the velocities and space positions of these objects from the initial velocity and spatial distributions detailed in Sec.~\ref{sec_spatial}. We draw the stochastic force $\gamma(t)$ 
independently at each time step and for each object from a Gaussian distribution of vanishing mean using as an underlying (pseudo)random number generator the Mersenne twister \cite{mersenne}. The variance of the Gaussian distribution is set by $\langle \gamma(t)\gamma(t')\rangle = 2\delta(t-t')$ in dimensionless units and discretized form. To this end, the correlation $\langle \gamma(t)\gamma(t')\rangle$ is integrated over one time step $dt$ and $\gamma(t)$ is assumed to be constant during that period. %

\begin{figure}
    \centering
    \includegraphics[width=\linewidth]{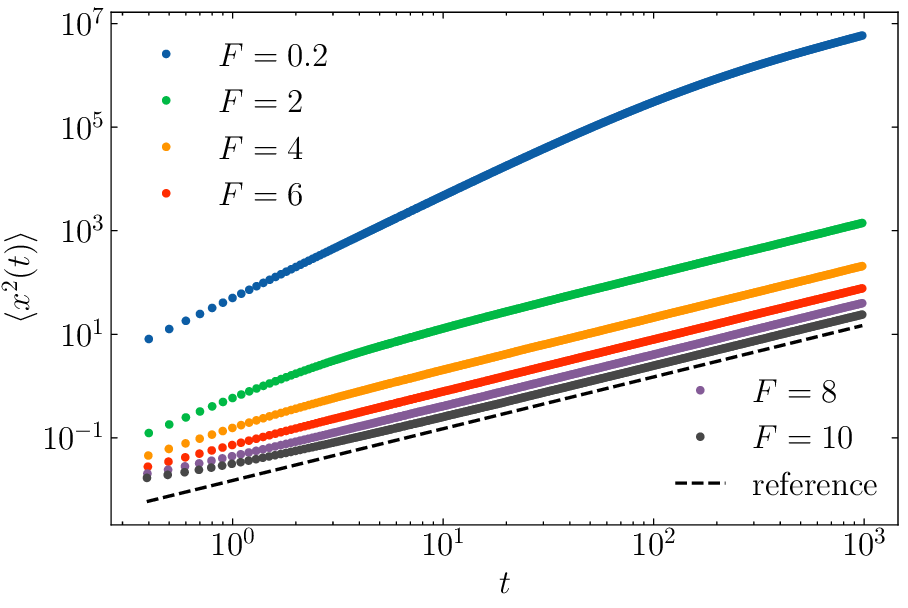}
    \caption{Mean-squared displacement $\langle x^2(t)\rangle$ over time on a double-logarithmic scale for different strengths $F$ of the Coulomb-tanh frictional force, as obtained from agent-based simulations of Eqs.~(\ref{eq:langevin}). Averages are obtained for simulations of $N=10^5$ objects over a total time $t_{max} = 1000$ with a discrete time step $dt= 10^{-3}$.
    As a visual guide, we include the black dashed line, which corresponds to a linear increase of $\langle x^2(t)\rangle\propto t$. We observe that stronger friction tends to support the approach of $\langle x^2(t)\rangle$ to this regime.
    \label{fig:msd}}
\end{figure}

Examples for $\langle x^2(t)\rangle$ over time are shown in Fig.~\ref{fig:msd} (see the caption for parameter settings in the simulations). 
First, in Fig.~\ref{fig:msd}, we address the same values of $F=0.2$ and $F=2$ as used for our illustrations in Fig.~\ref{fig:fvst_Ffric_comp}. We recall that the other values of $F$ listed in Fig.~\ref{fig:fvst_Ffric_comp} refer to the other types of considered friction for identical effective temperature $T_v$ of the moving objects. Additionally, we increase the strength of the frictional force up to $F=10$ to demonstrate the agreement between the results obtained via Eq.~(\ref{eq_D}) and the agent-based simulations.

The resulting diffusion coefficients for both approaches are compared in Tab.~\ref{tab:D_sim_theo}. We obtained the diffusion coefficients $D_{\text{sim}}$ from the simulations via linear fits to $\langle x^2(t)\rangle$ within the time interval $50<t<1000$, when the initial effects have decayed. The double integral for the analytical solution in Eq.~(\ref{eq_D}) was computed using \textit{Mathematica} \cite{Mathematica}. Deviations are smaller than 1~\%.

\begin{table}
\caption{\label{tab:D_sim_theo} Comparison of diffusion coefficients at different force parameters $F$ resulting in different temperatures $T_v$. The diffusion coefficients obtained by agent-based simulation $D_{\text{sim}}$ agree well with the theoretically predicted results $D$ from Eq.~(\ref{eq_D}).}
\begin{ruledtabular}
\begin{tabular}{ l | d d d d d }
 $F$ & 2 & 4 & 6 & 8 & 10 \\ 
 $D_{\text{sim}}$ & 0.717 & 0.106 & 0.0392 & 0.0203 & 0.0122 \\
 $D$ & 0.712 & 0.106 & 0.0393 & 0.0202 & 0.0123 \\
 $T_v$ & 0.823 & 0.323 & 0.198 & 0.142 & 0.111 \\
\end{tabular}
\end{ruledtabular}
\end{table}

\begin{figure}
    \centering
    \includegraphics[width=0.99\linewidth]{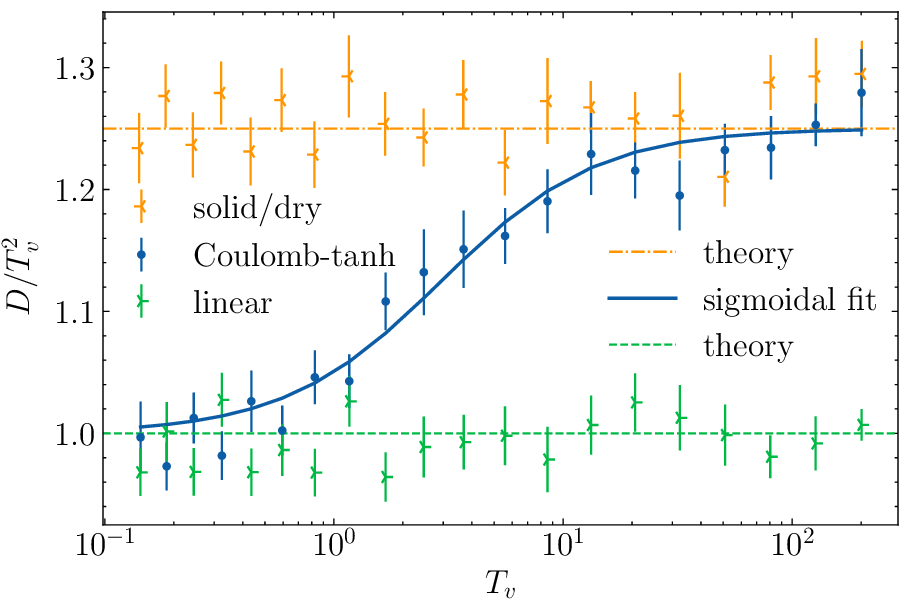}
    \caption{Diffusion coefficients for friction of the Coulomb-tanh type (blue), solid (dry, Coulomb) friction (orange), and linear friction (green) over a logarithmic temperature scale. The data were obtained using agent-based simulations for $N=10^4$ objects (time step of $dt=10^{-4}$). We determined the diffusion coefficients via linear fits to the mean-squared displacements in the time interval $50 < t < 10^4$. Error bars result from splitting the fitting interval into multiple time intervals and computing the standard deviation of the slopes of the mean squared displacement obtained in the different time intervals. Theoretical values for purely linear and purely solid (dry, Coulomb) friction are shown as dashed lines, see Appendix \ref{sec:app_D}. As a guide to the eye, the data for Coulomb-tanh friction were fitted by a sigmoidal function $\left[1 + \operatorname{sigm}(\alpha \ln( T_v/T_{c}))/4\right]T_v^2$ with fit parameters $\alpha$ and $T_c$. 
    }
    \label{fig:diff_coeffs}
\end{figure}

In the cases of linear and solid (dry, Coulomb) friction, we expect an initially quadratic dependence $D \propto T_v^2$ based on dimensional analysis. As explained above, friction of the Coulomb-tanh type interpolates between these two limiting cases when turning from small speeds or low $T_v$ to larger speeds or elevated $T_v$. Indeed, we find corresponding behavior, see Fig.~\ref{fig:diff_coeffs}. 

\section{Time evolution of the spatial statistics}\label{sec_spatial}

To obtain the spatial distribution function $f_x(x,t)$, it is necessary to solve the entire Fokker-Planck equation~(\ref{eq:FP}) for the complete distribution function $f(x,v,t)$. From there, we find $f_x(x,t)$ by integrating out the velocity component, $f_x(x,t)=\int_{-\infty}^{\infty}f(x,v,t)dv$. We performed this task numerically using \textit{Mathematica} \cite{Mathematica}. 

\begin{figure}
    \centering
    \includegraphics[width=0.99\linewidth]{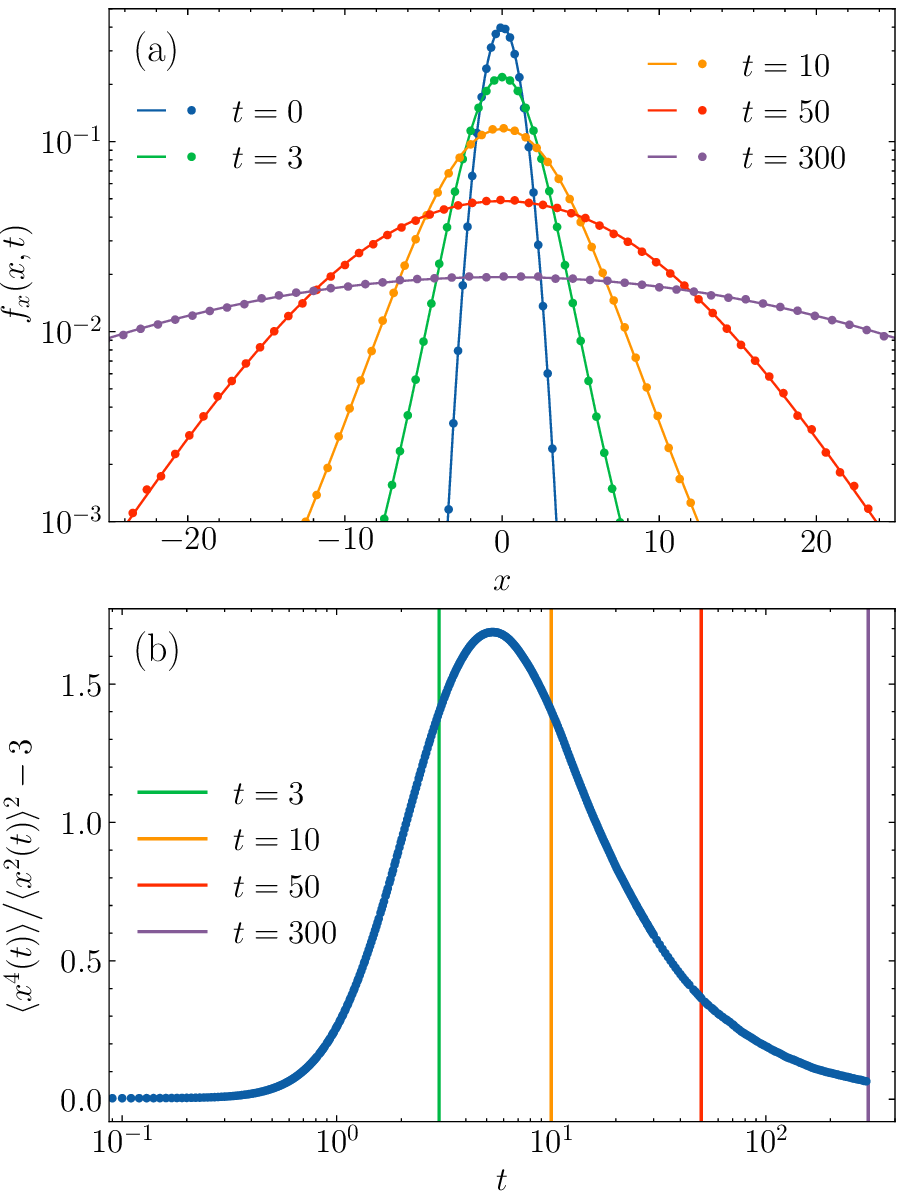}
    \caption{(a) Spatial distribution $f_x(x,t)$ at different times $t$, starting at time $t=0$ from a Gaussian distribution of standard deviation $\sigma_x = 1$ concerning the spatial component, multiplied by the stationary velocity distribution $f_{v,st}(v)$ as given by Eqs.~(4). The scale on the ordinate is logarithmic, and the strength of the Coulomb-tanh frictional force was set to $F=2$. Lines represent results as obtained by numerical solution of the Fokker-Planck equation, Eq.~(2), using \textit{Mathematica} \cite{Mathematica}. Dots mark results calculated through agent-based simulations of the Langevin equations~(\ref{eq:langevin}) from ensembles of $N=10^7$ objects (time step $dt = 10^{-3}$). 
    The spatial distributions in the agent-based simulations were determined using spatial bins of width $dx=0.1$. For better visibility, not all bin values are shown.
    (b) Time evolution of the excess kurtosis $\langle x^4(t)\rangle/\langle x^2(t)\rangle^2 - 3$ as extracted from the agent-based simulations, measuring the deviation of the kurtosis from that of a Gaussian distribution. Colored lines mark the instants of the snapshots in (a), except for time $t=0$. 
    Intermediate departure from the Gaussian value is largest around $t\approx 5$, when simultaneously we observe pronounced non-Gaussian tails in (a).
    }
    \label{fig:xdist}
\end{figure}

Apart from that, we again carried out agent-based simulations of the Langevin equations~(\ref{eq:langevin}) [see Sec.~\ref{sec_diff_ceoff}] now for an ensemble of $N=10^7$ objects. 
We compare the results obtained from the two methods, that is, from the solution of the Fokker-Planck equation~(\ref{eq:FP}) using \textit{Mathematica} \cite{Mathematica} and from the agent-based simulations of the Langevin equations~(\ref{eq:langevin}). To facilitate comparison between the results obtained by the two different approaches, we chose as an initial condition an identical product of a spatial distribution function and a velocity distribution function. The former was selected of Gaussian shape with standard deviation $\sigma_x=1$, and the latter as the stationary velocity distribution $f_{v,st}(v)$; see Eqs.~(\ref{eq:fvst}).
Spatial distributions $f_x(x,t)$ at different times $t$ as obtained from both methods are shown and compared with each other in Fig.~\ref{fig:xdist}(a). 
The results from the two approaches agree very well. Particularly, we observe pronounced non-Gaussian, rather exponential tails on intermediate time scales. %, here for $t\approx\tl{5}$. 
Those tails are absent in the initial condition of a narrow Gaussian form concerning the spatial component.  Thus, they are related and develop due to the action of the nonlinear Coulomb-tanh frictional force. The tails are pushed outward during the further course of time evolution. 

In addition, to further quantify the non-Gaussian character of the spatial distributions at intermediate timescales, we include in Fig.~\ref{fig:xdist}(b) the excess kurtosis $\langle x^4(t)\rangle/\langle x^2(t)\rangle^2 - 3$ as extracted from the agent-based simulations. It vanishes for spatial distributions of Gaussian form, here at $t=0$. In our case, it shows a maximum at around $t\approx5$. Afterward, it decreases again with proceeding time $t$. This temporal development is in line with our observations in Fig.~\ref{fig:xdist}(a).

\begin{figure}
    \centering
    \includegraphics[width=0.98\linewidth]{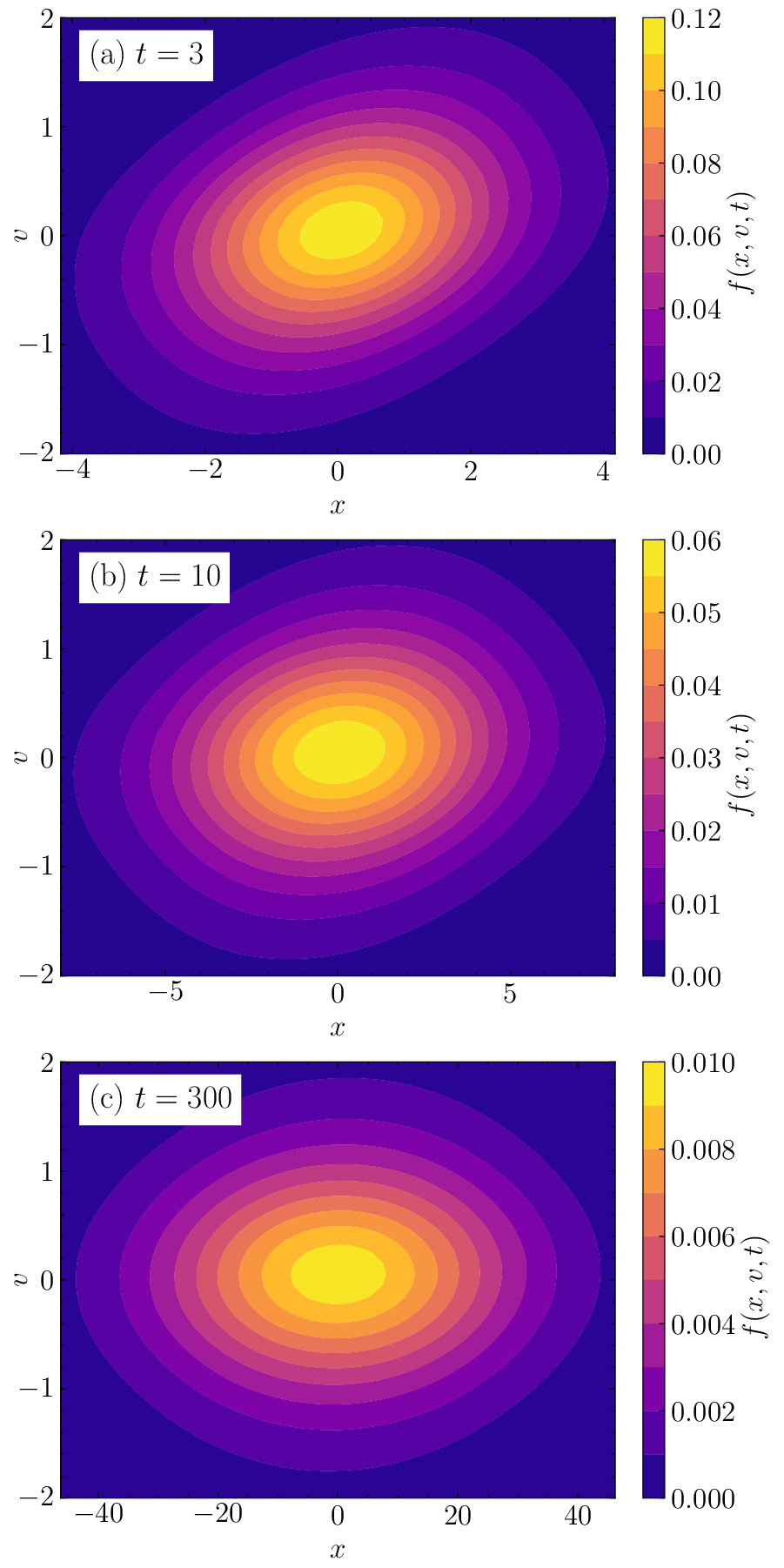}
    \caption{Color plots of the full probability distribution function $f(x,v,t)$ at different times (a) $t=3$, (b) $t=10$, and (c) $t=300$, as obtained by numerical solution of the Fokker-Planck equation~(\ref{eq:FP}), under the conditions already stated in the caption of Fig.~\ref{fig:xdist} using \textit{Mathematica} \cite{Mathematica}. 
    We stress the different scalings of both the abscissae and color scale bars due to the significant broadening of the distribution function along the space direction over time.
    \label{fig:complete_dist}}
\end{figure}

Furthermore, Fig.~\ref{fig:complete_dist} illustrates at some times the complete probability distribution function $f(x,v,t)$ as obtained by numerical solution of the Fokker-Planck equation, Eq.~(\ref{eq:FP}), using \textit{Mathematica} \cite{Mathematica}. At intermediate times, the distribution function becomes nonsymmetric with respect to the lines $x=0$ and $v=0$, see Fig.~\ref{fig:complete_dist}(a) and (b). Thus, spatial and velocity components couple to each other and the factorization into a space- and velocity-dependent contribution inherent to the initial condition is lost. At late times [see Fig.~\ref{fig:complete_dist}(c)] symmetry is roughly restored so that a factorization again becomes conceivable. 

\section{Conclusions}\label{sec_concl}

In summary, using statistical approaches, we investigated the stochastically driven motion of an object under Coulomb-tanh friction. This frictional force interpolates between linear friction at small speeds and solid (dry, Coulomb) friction at larger speeds, which is also reflected by the velocity distribution functions and diffusion coefficients for smaller and larger magnitudes of stochastic driving, respectively. 
Our specific motivation originates from stochastically driven objects that in addition to the driving mechanism, are subject to nonlinear frictional forces. Here, we consider the case of a frictional force that increases less substantially with speed at larger magnitudes of velocity. Such a situation occurs in scenarios of shear thinning. For instance, in that case, the environment may contain a surrounding shear-thinning viscous fluid or a subjacent shear-thinning layer deposited between an object and the supporting substrate. Depending on the specific characteristics of the employed materials, the situation would then to a certain degree be reflected by a frictional force of the Coulomb-tanh type.

It turns out that in one spatial dimension, the purely velocity-dependent part of the associated Fokker-Planck equation can be mapped onto a Schr\"odinger equation. There, the potential is of P\"oschl-Teller form. In this way, we were able to treat the velocity-dependent part of the statistical equation as well as the calculation of the diffusion coefficient analytically. Agent-based simulations were performed for comparison and confirmed our analytical results and numerical solution of the associated Fokker-Planck equation.

Our study can be extended in various directions. First, naturally, consequences for the physics in more than one spatial dimension should be evaluated. Here, we considered a one-dimensional situation, because our focus was on the possibility of an analytical solution and on pointing out the relation to the quantum-mechanical situation. 
In two (or even three) dimensions, the approach may still be valid, if the described dynamics are evaluated along the resulting one-dimensional trajectories. To this end, we assume a preferred body-fixed axis of the moving object, along which friction with the substrate is significantly lower than perpendicular to it \cite{menzel2022statistics}. Then, motion is mainly confined to directions along the present orientation of this axis. An additional equation of motion may apply to the dynamics of the axial orientation, yet it must not affect the translational dynamics along the one-dimensional trajectory for our description to remain valid.

Second, one could address the consequences of an additional linear friction term if added to the Langevin equation \cite{touchette2010brownian, menzel2011effect, menzel2022statistics}. Such a situation could result, for instance, if the motion of submillimeter particles takes place at the interface between two fluids, one of them viscous and the other one shear thinning. External stochastic driving could be imposed in an experiment, for instance, by optical tweezers or, for magnetic particles, by external magnetic field gradients. 

A specific example in this context is self-propelled particles or other objects that by themselves feature an independent active driving. Physical examples are hoppers on vibrating surfaces \cite{degennes2005brownian, narayan2007long, goohpattader2010diffusive, deseigne2010collective, scholz2018inertial, narayan2007long, lanoiselee2018statistical} or colloidal particles featuring a certain driving mechanism in liquid suspension \cite{jiang2010active, das2015boundaries, simmchen2016topographical, narinder2021active}. The dynamics of such objects have been addressed theoretically by setting the coefficient of the just-mentioned linear friction to negative values \cite{schweitzer1998complex, dunkel2001thermodynamics, lindner2008critical, romanczuk2012active, ohta2009deformable, hiraiwa2011dynamics, menzel2012soft}. Yet, higher-order nonlinear friction terms then are required for stabilization. Likewise, self-propulsion of individual objects is frequently introduced in theoretical studies by including an active forcing of constant magnitude \cite{ten2011brownian, pototsky2012active, ni2013pushing, menzel2015tuned, heidenreich2016hydrodynamic}. We can interpret this type of active driving as solid (dry, Coulomb) friction with a negative friction coefficient. Its consequences have been studied in detail \cite{menzel2022statistics}. In our situation, such a description would be stable, as long as the constant active driving is less in magnitude than the value at which the Coulomb-tanh friction term levels off, or if, again, a contribution of linear friction is added. 

\begin{acknowledgments}
T.L.\ thanks the Studienstiftung des Deutschen Volkes (German Academic Scholarship Foundation) for financial and academic support. 
A.M.M.\ acknowledges support by the Deutsche Forschungsgemeinschaft (German Research Foundation, DFG) through the Heisenberg Grant ME 3571/4-2. 
\end{acknowledgments}

\appendix

\section{Evaluation of the integral for normalization in Eq.~(\ref{eq:fvst_without_coefficient})} \label{sec_appA}

In this appendix, we solve the integral appearing in the denominator of Eq.~(\ref{eq:fvst_without_coefficient}), that is, we determine the normalization coefficient of the stationary velocity distribution function $f_{v,st}(v)$ leading to Eq.~(\ref{eq:fvst_with_coefficient}). 
To this end, for $F>0$, we define
\begin{equation} \label{eq:Z_definition}
    Z(F) :=\int_{-\infty}^{\infty} \frac{1}{\cosh^{F}(v)} dv.  \quad %\text{for}\quad F>0.
\end{equation}
First, we prove a recursion formula from $Z(F)$ to $Z(F+2)$ using integration by parts and trigonometric identities, 
\begin{eqnarray}
    Z(F+2) &=& \int_{-\infty}^{\infty} \frac{1}{\cosh^{F+2}(v)} dv \nonumber \\
    &=& {\left[ \frac{\tanh(v)}{\cosh^F(v)} \right]_{-\infty}^{\infty}} +F\int_{-\infty}^{\infty}\frac{\sinh^2(v)}{\cosh^{F+2}(v)} dv \nonumber \\
    &=& F\int_{-\infty}^{\infty}\frac{1}{\cosh^F(v)} dv \nonumber\\
    &&- F\int_{-\infty}^{\infty}\frac{1}{\cosh^{F+2}(v)} dv \nonumber\\
    &=& \frac{F}{F+1}\int_{-\infty}^{\infty}\frac{1}{\cosh^F(v)} dv .
\end{eqnarray}
Hence, we reach the recursion formula
\begin{equation} \label{eq:recZ}
    Z(F+2) = \frac{F}{F+1}Z(F) .
\end{equation}
We can evaluate the base case explicitly to $Z(2) = 2$.  

The same recursion relation applies to the expression 
\begin{equation}\label{eq:Z_expr}
    %Z(F) =
    2 \frac{2\cdot 4 \cdot \ldots \cdot (F-2)}{3 \cdot 5 \cdot \ldots \cdot (F-1)}.
\end{equation}
Moreover, it equally evaluates to $2$ when we set $F=2$. Therefore, the expression in Eq.~(\ref{eq:Z_expr}) can be identified with $Z(F)$ for all positive, even integer values of $F$. It can be rewritten as
\begin{equation}\label{eq:ZFprev}
    Z(F) = 2 \frac{(F-2)!!}{(F-1)!!} = \frac{2^{F+1}(F/2)!^2}{F!F},
\end{equation}
where $n!!$ denotes the double factorial, that is, the product of all positive integers of the same parity as $n$.

Yet, in our case, we remain with even-integer-valued parameters $F$, because in that case the Schr\"odinger equation~(\ref{eq:schroedinger}) is analytically solvable. Introducing the Gamma function $\Gamma$ in this situation, Eq.~(\ref{eq:ZFprev}) is identical to the expression
\begin{equation}
    Z(F) = \frac{\sqrt{\pi}\:\Gamma\!\left(\frac{F}{2}\right)}{\Gamma\!\left(\frac{F+1}{2}\right)}.
\end{equation}
This concludes our proof of Eq.~(\ref{eq:fvst_with_coefficient}) and its derivation from Eq.~(\ref{eq:fvst_without_coefficient}).

\section{Evaluation of $\langle v^2 \rangle_{st}$ to find Eq.~(\ref{eq:T_dimless})} \label{sec_appB}
Next, we compute the expression of the temperaturelike variable $T_v$ listed in Eq.~(\ref{eq:T_dimless}). It is given by the mean-squared velocity in the stationary state as introduced before this equation. %Eq.~(\ref{eq:T_dimless}). 

In parallel to the integral $Z(F)$ in Eq.~(\ref{eq:Z_definition}), we define
\begin{equation}
    S(F) := \int_{-\infty}^{\infty} \frac{v^2}{\cosh^F(v)}dv.
\end{equation}
It allows us to rewrite the expression for $T_v=\langle v^2\rangle_{st}$ as
\begin{equation} \label{eq:T_integral_definition}
    T_v(F) = \langle v^2\rangle_{st} = \int_{-\infty}^\infty v^2 f_{v,st}(v) \,dv = \frac{S(F)}{Z(F)}.
\end{equation}

Building on Appendix~\ref{sec_appA}, we first derive a recursion formula from $S(F)$ and $Z(F)$ to $S(F+2)$ using integration by parts and trigonometric identities,
\begin{eqnarray}
    S(F+2) &=& \int_{-\infty}^{\infty} \frac{v^2}{\cosh^{F+2}(v)}\, dv \nonumber \\
    &=& {\left[ \frac{v^2\tanh(v)}{\cosh^F(v)} \right]_{-\infty}^{\infty}} \nonumber \\
    &&{}- \int_{-\infty}^{\infty}\left(\frac{2v}{\cosh^{F}(v)}\right. \nonumber \\ 
    &&\qquad\left. {}-F \frac{v^2\sinh(v)}{\cosh^{F+1}(v)}\right)\tanh(v)\, dv \nonumber \\
    &=&{}-2\int_{-\infty}^{\infty}v\frac{\tanh(v)}{\cosh^F(v)}\, dv \nonumber \\
    && {}+ F\int_{-\infty}^{\infty}\frac{v^2(\cosh^2(v) - 1)}{\cosh^{F+2}(v)} dv \nonumber\\
    &=& 2\left[ \frac{v}{F\cosh^{F}(v)}\right]_{-\infty}^{\infty} - \frac{2}{F}\int_{-\infty}^{\infty}\frac{1}{\cosh^F{v}}\,dv \nonumber\\
    && {}+ F \int_{-\infty}^{\infty} \frac{v^2}{\cosh^{F}(v)} dv \nonumber \\ 
    && {}- F \int_{-\infty}^{\infty} \frac{v^2}{\cosh^{F+2}(v)}\, dv
    \nonumber\\
    &=& FS(F) - \frac{2}{F}Z(F) - FS(F+2) .
\end{eqnarray}    
% \am{Bitte Zeilen umbrechen, so dass in der Spalte bleiben; auch vor plus und minus in einer Klammer moeglich; $S(F+2)$ links weg in letzten beiden Zeilen.}
From there, it follows that
\begin{eqnarray}
     S(F+2) &=& \frac{F}{F+1}S(F) - \frac{2}{F(F+1)}Z(F).\label{eq:recursion_S}
\end{eqnarray}
Together with the recursion formula for $Z(F)$ in Eq.~(\ref{eq:recZ}), we deduce a recursion relation for $T_v$
\begin{eqnarray}
     T_v(F+2) &=& \frac{S(F+2)}{Z(F+2)} \nonumber \\ 
     &=&\frac{F}{F+1}\frac{S(F)}{Z(F+2)} - \frac{2Z(F)}{F(F+1)Z(F+2) \label{eq:recursion_T}}\nonumber\\
     &=& T_v(F) - \frac{2}{F^2}.
\end{eqnarray}
For $F=2$, the integral can be evaluated explicitly to $T_v(2) = \pi^2/12$, which would already suffice to inductively deduce the value $T_v(F)$ for all even-integer-valued parameters $F$.

We note that the same recursion relation as in Eq.~(\ref{eq:recursion_T}) follows, if we define
\begin{equation} \label{eq:def-Tv-psi1}
    T_v(F) := \frac{1}{2} \psi^{(1)}\left(\frac{F}{2}\right)
\end{equation}
for positive values of $F > 0$. More precisely, we remain in our analytical considerations with positive, even integer values of $F$, as explained above. $\psi^{(1)}$ is the second polygamma function, also known as the trigamma function. This identification also reproduces the base case $T_v(2) = \pi^2/12$. Therefore, Eq.~(\ref{eq:def-Tv-psi1}) gives the correct expression in our description. Thus, we have demonstrated the validity of Eq.~(\ref{eq:T_dimless}).

\section{Normalization of the scattering states in Eq.~(\ref{eq:eigenfunction_free})} \label{sec_scatt_states}

Here, we explain how to find the expressions of the normalized scattering states $\psi_{ik}(v)$ in Eq.~(\ref{eq:eigenfunction_free}). 
We rewrite the associated Legendre functions by means of the hypergeometric function $_2F_1$, 
\begin{eqnarray}
    P^{ik}_l(\tanh v) &=& \frac{1}{\Gamma(1-ik)}\left[\frac{1+\tanh(v)}{1-\tanh(v)} \right]^{ik/2} \nonumber\\ && \times {}_2F_1\left(-l, l+1, 1-ik, \frac{1-\tanh(v)}{2}\right) .\nonumber \\ &&
\end{eqnarray}
Next, we consider its limit for $v \to \infty$, where we find
\begin{eqnarray}
\lefteqn{\lim_{v\rightarrow\infty}P^{ik}_l(\tanh v)}\qquad\qquad
\nonumber
\\
    &=& \frac{1}{\Gamma(1-ik)}e^{ikv}\, {}_2F_1(-l, l+1, 1-ik, 0) \nonumber \\
    %&=& \frac{1}{\Gamma(1-ik)}e^{ikv} \sum_{n=0}^{\infty} \. frac{(-l)_n(l+1)_n}{(1-ik)_n} \frac{0^n}{n!} \nonumber \\
    &=&  \frac{1}{\Gamma(1-ik)}e^{ikv} ,
    \label{eq:limit_norm}
\end{eqnarray}
where $\Gamma$ denotes the gamma function. %and  $(q)_n$ the rising Pochhammer symbol. The latter is defined as
%\begin{equation}
%    (q)_n = \begin{cases} 1 & n=0\\ q(q+1)\dots(q+n-1)(q+n) & n > 0 \end{cases},
%\end{equation} which results in $(q)_0 = 1$. 
Thus, we obtain a complex exponential scaled by a factor of $\Gamma(1-ik)^{-1}$.

The P\"oschl-Teller potential has the particular property of being reflectionless \cite{lekner2007reflectionless}. Therefore, the same amplitude is expected for $v\to - \infty$. % also consists of a right traveling wave with the same amplitude. 
If we normalize the scattering states in a large box %of size $V$ 
and increase the size of the box, the contribution of the central part, where the potential is most pronounced, vanishes. We thus use the normalization factor of free particles. Consequently, in one spatial dimension, we multiply Eq.~(\ref{eq:limit_norm}) by $\Gamma(1-ik)(2\pi)^{-1/2}$.

\section{Diffusion coefficient for solid (dry, Coulomb) friction} \label{sec:app_D}
Finally, we revisit the derivation of the diffusion coefficient in the case of pure solid (dry, Coulomb) friction to verify the results of our simulations, as shown in Fig.~\ref{fig:diff_coeffs}. In this case, the Langevin equations, instead of Eqs.~(\ref{eq:langevin}), read \cite{degennes2005brownian}
\begin{subequations}
    \begin{eqnarray}
        \frac{dv}{dt} &=& {}- F\sigma(v) + \gamma(t),\\
        \frac{dx}{dt} &=& v,
    \end{eqnarray}
\end{subequations}
% \am{Punkte bitte auch als Ableitungen}
where we rescaled all variables to dimensionless quantities as in Sec.~\ref{sec_langevin}. Here, $F>0$ represents the strength of the frictional force, and $\sigma(v) = |v|/v$ with $\sigma(0) = 0$ is the sign function. $\gamma(t)$ includes a Gaussian, white stochastic force of zero mean and correlation in time $\langle \gamma(t)\gamma(t')\rangle = 2\delta(t-t')$. Instead of Eq.~(\ref{eq:FP}), this leads us to the Fokker-Planck equation
\begin{equation}
    \partial_t f(x,v,t) = \Bigl\{\!{} - v \partial_x + \partial_v F \sigma(v) + \partial_v^2\Bigr\} \,f(x,v,t).
\end{equation}
We integrate this equation over space to obtain
\begin{equation}\label{eq:FP_v_solid_dry}
    \partial_t f_v(v,t) = \partial_v\Bigl\{F\sigma(v) +\partial_v\Bigr\} f_v(v,t).
\end{equation}
Setting $\partial_t f_v(v,t)=0$, we find the stationary velocity distribution $f_{v,st}(v) = F\exp(-F|v|)/2$. 

Next, we introduce $g_v(v,t) = f_{v,st}^{-1/2}(v) f_v(v,t)$. From  Eq.~(\ref{eq:FP_v_solid_dry}), we obtain, for $g_v(v,t)$, the equation
\begin{equation}
    \partial_t g_v(v,t) = \Biggl\{-\frac{F^2}{4}+ F\delta(v) + \partial_v^2 \Biggr\} g_v(v,t),
\end{equation}
where $\delta(v)$ is the Dirac delta function. The operator on the right-hand side is Hermitian, and we perform spectral decomposition. 

Introducing $g_v(v,t)=\sum_k\psi_k(v)\exp(-E_kt)$, 
the only bound eigenfunction $\psi_0(v) = f_{v,st}^{1/2}(v)$ is associated with the eigenvalue $E_0 = 0$. Furthermore, there is a continuum of scattering states that can be split into even ($e$) and odd ($o$) eigenfunctions for $k>0$. These eigenfunctions, respectively, read
\begin{eqnarray}
    \psi_{k,e}(v) &=& \frac{1}{\sqrt{\pi}}\cos\left(k|v| + \arctan\left(\frac{F}{2k}\right)\right), \\
    \psi_{k,o}(v) &=& \frac{1}{\sqrt{\pi}}\sin(kv).
    \label{eq:eigenfunc_odd_solid_dry}
\end{eqnarray}
They are associated with the eigenvalues $E_k = F^2/4 + k^2$. % \am{bitte Vorzeichen checken}. 

Eventually, the diffusion coefficient is calculated in analogy to Eq.~(\ref{eq_D}). Only the odd eigenfunctions from Eq.~(\ref{eq:eigenfunc_odd_solid_dry}) contribute due to symmetry, so that after integration over time we obtain
\begin{eqnarray}
    D &=& \int_{0}^{\infty} \frac{\left|\langle \psi_0|v|\psi_{k,o}\rangle\right|^2}{E_k}\, dk = \frac{\pi F^3}{2} \int_0^{\infty} \frac{k^2 \; dk}{\left(\frac{F^2}{4} + k^2 \right)^5} \nonumber \\
    &=& \frac{5}{F^4}. 
\end{eqnarray}
Besides, direct calculation leads to $T_v = \langle v^2\rangle_{st} = 2/F^2$. 
Therefore, 
\begin{equation}
    D = \frac{5}{4}T_v^2.
\end{equation}
The results from our simulations presented in Fig.~\ref{fig:diff_coeffs} confirm this relation.

%\section*{References}
%\appendix

%\vspace{-.7cm}

%\bibliography{references}


\end{document}